\documentclass[10pt]{algotel}

\usepackage[utf8]{inputenc}
\usepackage[T1]{fontenc}
\usepackage[french]{babel}
\usepackage{color}
\usepackage{xspace}
\usepackage{tikz}

\usepackage[lined,boxed,noend]{algorithm2e} 
\usepackage{amsmath,amsfonts,amssymb,amsthm}
\usepackage{multicol}
\usepackage{multirow}

\newtheorem{thm}{Theorem} 
\newtheorem{lem}{Lemma} 

\newtheorem*{Proof}{\textmd{\textit{Démonstration}}}

\newcommand{\G}{\mathcal{G}}

\renewcommand{\k}{\delta}
\newcommand{\Pred}{\mathcal{P}}
\newcommand{\PredAdd}{\mathcal{P}^{+}}
\newcommand{\FTS}{\ensuremath{\G^*_{st}}\xspace}
\newcommand{\FT}{\ensuremath{\G^*}\xspace}
\newcommand{\leadstoo}{\overset{st}{\leadsto}}
\newcommand{\TVG}{\ensuremath{{\G=(V,E,\mathcal{T},\rho,\zeta)}}\xspace}

\newcommand{\fixme}[1]{{\color{red}{(fixme)}}\xspace}

\title{Un algorithme de test pour la connexité temporelle des graphes dynamiques de faible densité\thanks{Une version longue en anglais est disponible sur arXiv~\cite{BCCJN14e}. Ce travail est partiellement subventionné par la DGA via une bourse the thèse (n$^\circ$ 2013 60 0074).}}

\author{Matthieu Barjon
  \and Arnaud Casteigts
  \and Serge Chaumette\\
  \and Colette Johnen
  \and Yessin M. Neggaz
}

  \address{LaBRI, Université de Bordeaux
  }

\begin{document}
\maketitle

\begin{abstract}
Nous considérons le problème de tester si un graphe dynamique donné est temporellement connexe, {\it i.e.} s'il existe un chemin temporel (aussi appelé {\em trajet}) entre toute paire de sommets. Nous considérons une version simplifiée du problème où la dynamique est représentée par un graphe évolutif non-temporisé $\G=\{G_1,G_2,...,G_{\k}\}$ dont l'ensemble des sommets est invariant et les arêtes sont orientées (arcs). Deux variantes du problème sont étudiées, selon que l'on autorise la traversée consécutive d'un seul ou d'un nombre illimité d'arcs à chaque étape (trajets stricts {\it vs} non-stricts). 

Dans le cas des trajets stricts, deux algorithmes pré-existants pour d'autres problèmes peuvent être adaptés. Cependant, nous montrons qu'une approche dédiée permet d'obtenir une meilleure complexité en temps que le premier algorithme dans tous les cas, et que le second dans certaines familles de graphes, notamment les graphes dont la densité est faible à tout instant (bien que potentiellement élevée à travers le temps). La complexité de notre algorithme est en $O(\k\mu n)$, où $\k$ est le nombre d'{\em étapes} $|\G|$ et $\mu=max(|E_i|)$ est le nombre maximal d'arcs pouvant exister à un instant donné. Ce paramètre est à contraster avec $m=|\cup E_i|$, l'union de tous les arcs apparaissant au cours du temps. En effet, il n'est pas rare qu'un scénario de mobilité exhibe à la fois un $\mu$ petit et un $m$ grand. Nous caractérisons les principales valeurs charnières de $\k, \mu$ et $m$ permettant de décider quel algorithme utiliser.
Dans le cas des trajets non-stricts, pour lesquels nous ne connaissons pas d'algorithme existant, nous montrons que notre algorithme peut être adapté pour répondre à la question, et ce, toujours en $O(\k\mu n)$.

Nos deux algorithmes construisent graduellement la fermeture transitive des trajets stricts (notée $\G^*_{st}$) ou non-stricts (notée $\G^*$) à mesure que les arcs sont examinés. Ce sont des algorithmes de type {\em streaming} qui sont aussi capables d'arrêter leur exécution sitôt la connexité temporelle atteinte. Un sous-produit intéressant est de rendre $\G^*_{st}$ et $\G^*$ disponibles pour de futures requêtes d'accessibilité temporelle de type source-destination.
\end{abstract}

\section{Introduction}
Les appareils connectés et mobiles tels que les téléphones portables, satellites, voitures ou robots forment des réseaux très dynamiques où la connexité entre n\oe uds évolue rapidement et continuellement. De plus, la topologie du réseau à un instant donné n'est généralement pas connexe, voire même très peu dense dans certains scenarios. Cependant, même dans ces cas extrèmes, une autre forme de connexité s'établit à travers le temps et l'espace, par le biais de communications tolérantes aux délais (mécanismes de type \og{\em store-carry-forward}\fg). On parle alors de {\em connexité temporelle}.

Nous nous intéressons au problème de tester automatiquement si un graphe dynamique donné est temporellement connexe. Autrement dit, déterminer s'il existe un chemin temporel ({\em journey} en anglais, {\em trajet} en français) entre toute paire de n\oe uds dans le réseau. Une notion clé est celle de fermeture transitive des trajets, introduite dans~\cite{BF03}. Il s'agit d'un graphe statique orienté (même si le graphe dynamique est non-orienté) dont les arcs représentent les possibilités de trajets. De cette structure peut être déduite l'appartenance d'un graphe dynamique à plusieurs familles de graphes~\cite{CCF09}, en particulier la famille des graphes temporellement connexes ({\it i.e.,} celle dont la fermeture transitive est un graphe complet). Nous nous intéressons à deux variantes: fermeture transitive stricte (\FTS) ou non-stricte (\FT), selon que l'on autorise la traversée consécutive d'un seul arc ou d'un nombre d'arcs illimité à chaque étape ({\it i.e.} trajets stricts {\it vs.} non-stricts).

Dans le cas des trajets stricts, plusieurs algorithmes peuvent être adaptés pour calculer \FTS. Trois de ces algorithmes sont proposés dans~\cite{BFJ03}, chacun permettant de calculer les trajets optimaux d'un sommet vers tous les autres selon un critère donné (au plus tôt, au plus court, au plus rapide). N'importe lequel peut être adapté au calcul de \FTS.  Le plus rapide des trois (trajets au plus tôt) a un temps d'exécution en $O(m\log \k + n\log n)$, d'où un temps total de $O(n(m\log \k + n\log n))$ pour tester les trajets depuis chaque sommet.

\label{sec:paris}
Un autre algorithme, calculant une généralisation de la fermeture transitive des trajets, a été proposé dans~\cite{WDCG12}. Cette généralisation, appelée {\em graphe d'accessibilité dynamique}, correspond à une fermeture transitive des trajets paramétrée par une date de départ et une durée maximale pour les trajets, ainsi qu'un délai de traversée d'arête. Il s'applique à des graphes dynamiques donnés sous la forme de TVG~\cite{CFQS12} ({\em time-varying graphs}), à savoir un quintuplet $\TVG$ où $\mathcal{T}$ est le domaine temporel (en l'occurrence $\mathbb{R}^+$) et $\rho$ et $\zeta$ sont des fonctions qui renseignent sur la présence et la latence d'une arête donnée à un instant donné. L'algorithme proposé peut également être utilisé pour calculer $\G^*_{st}$. La complexité de cet algorithme est en $O(\k \log \k\ mn\log n)$.

Nous proposons une approche dédiée au calcul de la fermeture transitive (d'abord stricte) d'un graphe évolutif non-temporisé orienté $\G=\{(V,E_i)\}$ qui permet d'obtenir une meilleure complexité en temps que l'adaptation de~\cite{WDCG12} dans tous les cas, et que l'adaptation de~\cite{BFJ03} pour une large famille de graphes dynamiques, en particulier ceux dont la densité est faible à tout instant, bien qu'arbitrairement dense à travers le temps. La complexité de notre algorithme est en $O(\k \mu n)$, où $\k=|\G|$ est le nombre d'étapes dans $\G$ et $\mu=max(|E_i|)$ est le nombre maximal d'arcs pouvant exister simultanément. Comme évoqué dans le résumé, ce dernier paramètre est à contraster avec $m=|\cup E_i|$, le nombre total d'arcs pouvant exister au cours du temps, l'écart entre les deux pouvant être très grand.
Dans le cas des trajets non-stricts, pour lequel nous ne connaissons pas d'algorithme existant, nous montrons que l'algorithme que nous proposons peut être adapté directement pour répondre à la question, et ce, toujours en $O(\k\mu n)$. Cette variation repose sur une double fermeture transitive, l'une relative aux étapes de $\G$ (comme dans le cas des trajets stricts), l'autre relative aux arcs dans un $G_i$ donné. Autrement dit, l'une est de nature temporelle, l'autre de nature statique, les deux étant combinées pour aboutir au résultat.

Nos deux algorithmes sont de type {\em streaming} et sont capables d'arrêter leur exécution sitôt la connexité temporelle atteinte. Un sous-produit de l'exécution est de rendre $\G^*_{st}$ et $\G^*$ disponibles pour d'éventuelles requêtes ultérieures de type {\em st-}connexité (temporelle), qui se réduisent alors à de simples requêtes d'incidence dans un graphe statique.


\section{Modèle et notations}
\label{sec:model}

Soit $\G$ un graphe évolutif non-temporisé orienté $\{G_i=(V,E_i)\}$. Il existe un trajet {\em non-strict} de $u$ vers $v$ dans $\G$ si et seulement si il existe une suite d'arcs $e_1, e_2, ..., e_p$ reliant $u$ à $v$ telle que pour tout $j\in 1..p$$-$$1$, $e_{j}\in E_i \implies \exists E_{i'\ge i}, e_{j+1}\in E_{i'}$. Si l'inégalité $i'>i$ est stricte, on parle de trajet {\em strict}. L'existence d'un trajet non-strict (resp. strict) de $u$ vers $v$, lorsque le contexte est implicite, est notée $u\leadsto v$ (resp. $u\leadstoo v$) sans préciser le graphe $\G$. Ainsi, dans un trajet strict, au plus un arc peut être traversé durant une même étape $i$, tandis que dans un trajet non-strict, le nombre d'arêtes pouvant être traversées lors d'une étape est illimité.

La fermeture transitive (non-stricte) d'un graphe dynamique $\G$ est le graphe {\em statique} orienté $\FT=(V,E^*)$ tel que $(u,v)\in E^* \Leftrightarrow u\leadsto v$. La fermeture transitive stricte de $\G$ est le graphe statique orienté $\FTS=(V,E_{st}^*)$ tel que $(u,v)\in E_{st}^* \Leftrightarrow u\leadstoo v$.
Notez que $\G^*$ est orienté quelle que soit la nature (orientée ou non) des arêtes de $\G$, car la dimension temporelle induit sa propre orientation.

Étant donné $\G$, on note $\k=|\G|$ le nombre d'étapes dans $\G$. On distingue deux paramètres pour rendre compte du nombre d'arcs dans le graphe : le nombre maximal d'arcs existant à une même étape, i.e. $\mu=max(|E_i|)$, et le nombre total d'arcs pouvant exister au cours du temps, i.e. $m=|\cup E_i|$. Bien sûr, quelque soit le graphe considéré, on a $m\ge \mu$, et même souvent $m \gg \mu$.

\section{Calcul de la fermeture transitive des trajets stricts}
\label{sec:strict}

Nous proposons ci-dessous un algorithme de calcul de la fermeture transitive stricte $\FTS$ dans le cas général où $\G$ est orienté. Le principe de l'algorithme est de construire, étape après étape, la liste de tous les prédécesseurs de chaque sommet, i.e., pour un sommet $v$, l'ensemble $\{u : u\leadstoo v\}$. Soit $\Pred(v,t)$ l'ensemble des prédecesseurs de $v$ à l'issue des $t$ premières étapes (i.e. en tenant compte des ensembles d'arêtes $E_1, ..., E_t$). A l'étape $i$, le c\oe ur du traitement consiste à ajouter $\Pred(u,i-1)$ à $\Pred(v,i)$ pour chaque arête $(u,v)\in E_i$. En pratique, seules deux variables $\Pred(v)$ et $\PredAdd(v)$ sont maintenues pour chaque n\oe ud $v$, où $\PredAdd(v)$ contient les nouveaux prédécesseurs de v (ajoutés durant l'étape courante). Le détail des traitements est donné par l'Algorithme~1 dans la version longue du papier~\cite{BCCJN14e}.


\begin{lem}
  \label{lem:nbpred}
  Pour tout $v \in V$, $|\Pred(v)|\le \k\mu$, i.e., un noeud ne peut avoir plus de $\k\mu$ prédécesseurs.
\end{lem}

\begin{Proof}[\textit{par l'absurde}]
  S'il existe un n\oe ud $v$ tel que $|\Pred(v)\setminus v|> \k\mu$, alors, par définition, il existe plus de $\k\mu$ sommets $u$ différents de $v$ tels que $u\leadsto v$. Chacun de ces sommets est donc l'origine d'au moins un arc, ce qui implique que plus de $\k\mu$ arcs distincts ont existé.\qed
\end{Proof}

\begin{thm}
  L'Algorithme 1 calculant la fermeture transitive stricte d'un graphe $\G$ a une complexité en temps en $O(\k \mu n)$.
\end{thm}

\begin{Proof}
La boucle d'initialisation est linéaire en $n$. Vient ensuite la boucle principale, qui itère autant de fois qu'il y a d'étapes dans $\G$, i.e. $\k$ fois. Elle comporte trois sous-boucles, chacune étant dominée par $O(|E_i|\cdot n)=O(\mu n)$. Enfin, la construction de la fermeture transitive, si cette dernière n'est pas complète prématurément, consiste en une boucle qui, pour chaque n\oe ud, itère sur ses prédécesseurs. Or, on sait que le nombre de prédecesseur d'un n\oe ud donné ne peut excéder $\k\mu$ (Lemme~\ref{lem:nbpred}). Cette dernière boucle est donc elle aussi contenue dans $O(\k\mu n)$.\qed
\end{Proof}

\section{Calcul de la fermeture transitive des trajets non-stricts}
\label{sec:non-strict}

Dans cette section, nous nous intéressons au calcul de $\FT$, i.e. la fermeture transitive des trajets où un nombre illimité d'arêtes peut être traversé à chaque étape (trajets non-stricts). Une simple observation nous permet de réutiliser l'Algorithme~1 de manière quasiment directe. En effet, la relaxation de la contrainte que les trajets sont stricts implique qu'à chaque étape $i$, si un chemin (au sens classique) existe de $u$ vers $v$, alors $u$ peut joindre $v$ à cette même étape. L'algorithme consiste donc à pré-calculer, à chaque étape, la fermeture transitive (au sens classique, statique du terme) des arcs présents dans $G_i$, résultant en un graphe $G_i^*$ dont les arcs correspondent aux chemins dans $G_i$. L'Algorithme~1, appliqué ensuite au graphe dynamique $\{G_i^*\}$, produit ainsi la fermeture transitive $\FT$ des trajets {\em non-stricts} de $\G$.\\

La complexité en temps de cet algorithme dépend essentiellement du coût requis pour calculer la fermeture transitive statique $G_i^*$ des graphes $G_i$. Cela peut être fait par une recherche en profondeur (DFS) ou en largeur (BFS), exécutée  depuis chaque sommet dans $G_i$, chacune de ces exécution ayant un coût en $O(|E_i|)=O(\mu)$. Ainsi, le surcoût engendré par ce traitement reste confiné dans le même ordre de grandeur que nous avons identifié précédemment, à savoir $O(\k \mu n)$.

\section{Comparaison}
\label{sec:comparaison}

Cette section compare la complexité de notre algorithme à celle de la stratégie utilisant le calcul des trajets au plus tôt de~\cite{BFJ03}. Cette stratégie, qui revient à exécuter l'algorithme depuis chaque sommet, a une complexité totale en $O(n(m\log \k + n\log n))$, où $m$ est le nombre total d'arêtes pouvant exister au cours du temps, i.e. $|\cup E_i|$, et non $\mu$. 

Il s'agit donc de comparer cet ordre de grandeur à $O(\k\mu n)$, ou après simplification par $n$, de comparer $O(\k\mu)$ à $O(m\log \k + n\log n)$. Ces grandeurs appartiennent à un espace à quatre dimensions : $\mu, m, \k$ et $n$; il n'est donc pas aisé de les comparer. Nous proposons de les étudier asymptotiquement en $n$, en faisant varier les rapports entre $\mu, m$ et $\k$. Précisément, nous faisons varier les ordres de grandeur de $\mu$ et $m$ (densité \og instantanée\fg\  {\it vs.} densité \og cumulée\fg) pour plusieurs ratios de valeurs possibles entre $\k$ et $n$ (i.e. nombre d'étapes dans $\G$ en fonction de $n$). Le tableau proposé (Table~\ref{fig:tableau}) contient 60 résultats, dont une dizaine mettent en évidence là ou a lieu le basculement entre les deux algorithmes. Pour simplifier la vérification de ces résultats, nous fournissons dans la colonne de droite une expression intermédiaire, obtenue après simple substitution de $\mu$ et $m$ dans les deux expressions à comparer.

\begin{table}[h]
  \centering
\begin{tabular}{|@{\,}c@{\,}|@{\,}c@{\,}||@{\,}c@{\,}|@{\,}c@{\,}|@{\,}c@{\,}|@{\,}c@{\,}|@{\,}c@{\,}||@{\,}c@{\,}|}
  \hline
  \multirow{2}{*}{$\mu=\Theta(.)$}&\multirow{2}{*}{$m=\Theta(.)$}&\multirow{2}{*}{$\k=\Theta(\log n)$}&\multirow{2}{*}{$\k=\Theta(\sqrt n)$}&\multirow{2}{*}{$\k=\Theta(n)$}&\multirow{2}{*}{$\k=\Theta(n^2)$}&\multirow{2}{*}{$\k=\Theta(e^n)$}&Calcul intermédiaire\\
  &~&~&~&~&~&~&$\Theta(.) \pm \Theta(.)$\\\hline
$\log n$&$n^2$&$-$&$-$&$-$&$\approx$&$+$&$\k\log n \pm n^2\log \k$\\
$\sqrt n$&$n^2$&$-$&$-$&$-$&$+$&$+$&$\k\sqrt n \pm n^2\log \k$\\
$n$&$n^2$&$-$&$-$&$-$&$+$&$+$&$\k n \pm n^2\log \k$\\
$n\log n$&$n^2$&$-$&$-$&$\approx$&$+$&$+$&$\k(n\log n) \pm n^2\log \k$\\
$n^2$&$n^2$&$+$&$+$&$+$&$+$&$+$&$\k n^2 \pm n^2\log \k$\\\hline
$\log n$&$n\log n$&$-$&$-$&$-$&$+$&$+$&$\k\log n \pm (n\log n)\log \k$\\
$\sqrt n$&$n\log n$&$-$&$-$&$+$&$+$&$+$&$\k\sqrt n \pm (n\log n)\log \k$\\
$n$&$n\log n$&$-$&$+$&$+$&$+$&$+$&$\k n \pm (n\log n)\log \k$\\
$n\log n$&$n\log n$&$+$&$+$&$+$&$+$&$+$&$\k \pm \log \k$\\\hline
$\log n$&$n$&$-$&$-$&$\approx$&$+$&$+$&$\k\log n \pm n\log \k+n\log n$\\
$\sqrt n$&$n$&$-$&$-$&$+$&$+$&$+$&$\k\sqrt n \pm n\log \k+n\log n$\\
$n$&$n$&$\approx$&$+$&$+$&$+$&$+$&$\k n \pm n\log \k+n\log n$\\\hline
\end{tabular}
\caption{\label{fig:tableau}Comparaison de la complexité en temps de notre algorithme à l'adaptation de l'algorithme de~\cite{BFJ03}. {\it Les cases $-$ (resp $+$, $\approx$) indiquent les plages de paramètres pour lesquelles notre solution a une complexité asymptotique plus faible (resp plus forte, du même ordre de grandeur)}.}
\end{table}

En résumé, le tableau confirme que notre solution se comporte d'autant mieux que l'écart entre densité \og instantanée\fg et densité \og cumulée\fg est élevé, ce qui n'est pas surprenant. Il n'est pas surprenant non plus, au vu de la présence des facteurs $\k$ versus $\log \k$, que notre solution soit plus efficace lorsque le nombre d'étapes est relativement faible. En outre, le tableau révèle plusieurs éventails de valeurs naturelles où notre solution se comporte mieux, comme par exemple pour les trios $(\mu, m, \k)$ vallant $(O(n),\Theta(n^2),O(n))$, ou $(O(\log n),\Omega(n\log n),O(n))$, ou bien $(O(\log n), \Omega(n), o(n))$, ou encore $(O(\sqrt n)), \Omega(n), O(\sqrt n))$.

Enfin, nous pensons que l'impact coûteux du paramètre $\k$ dans la complexité théorique de notre algorithme doit être relativisé, eu égard au fait que l'algorithme termine dès que la connexité temporelle est atteinte. En effet, si l'on considère des modèles de graphes dynamiques aléatoires tels que les graphes à évolution arête-markovienne~({\em Edge-Markovian Evolving Graphs}), la connexité temporelle s'établit avec forte probabilité après un nombre sous-logarithmique d'étapes. La performance de notre algorithme dans les scénarios représentés par ce type de modèle correspondrait donc, en réalité, à la colonne la plus à gauche du tableau.

\end{document}